# Feeling Machines: Ethics, Culture, and the Rise of Emotional AI

**Authors**: Vivek Chavan *(Fraunhofer IPK & TU Berlin)*, Arsen Cenaj *(Université Sorbonne Paris Nord)*, Shuyuan Shen *(University of Augsburg)*, Ariane Bar *(EM Lyon Business School)*, Srishti Binwani *(EPITA, Paris)*, Tommaso Del Becaro *(Università di Pisa)*, Marius Funk *(University of Augsburg)*, Lynn Greschner *(University of Bamberg)*, Roberto Hung *(Universidad Central de Venezuela)*, Stina Klein *(University of Augsburg)*, Romina Kleiner *(RWTH Aachen University)*, Stefanie Krause *(Harz University of Applied Sciences)*, Sylwia Olbrych *(RWTH Aachen University)*, Vishvapalsinhji Parmar *(University of Passau)*, Jaleh Sarafraz *(SCAI, Sorbonne Université)*, Daria Soroko *(University of Hamburg)*, Daksitha Withanage Don *(University of Augsburg)*, Chang Zhou *(University of Augsburg)*, Hoang Thuy Duong Vu *(ICM, LIP6 UMR 7606 CNRS)*, Parastoo Semnani *(TU Berlin)*, Daniel Weinhardt *(Osnabrueck University)*, Elisabeth Andre *(University of Augsburg)*, Jörg Krüger *(TU Berlin)*, Xavier Fresquet *(Sorbonne Université)*

**Abstract**[1]

This paper explores the growing presence of emotionally responsive artificial intelligence through a critical and interdisciplinary lens. Bringing together the voices of early-career researchers from multiple fields, it explores how AI systems that simulate or interpret human emotions are reshaping our interactions in areas such as education, healthcare, mental health, caregiving, and digital life.

The analysis is structured around four central themes: the ethical implications of emotional AI, the cultural dynamics of human-machine interaction, the risks and opportunities for vulnerable populations, and the emerging regulatory, design, and technical considerations. The authors highlight the potential of affective AI to support mental well-being, enhance learning, and reduce loneliness, as well as the risks of emotional manipulation, over-reliance, misrepresentation, and cultural bias.

Key challenges include simulating empathy without genuine understanding, encoding dominant sociocultural norms into AI systems, and insufficient safeguards for individuals in sensitive or high-risk contexts. Special attention is given to children, elderly users, and individuals with mental health challenges, who may interact with AI in emotionally significant ways. However, there remains a lack of cognitive or legal protections which are necessary to navigate such engagements safely.

The report concludes with ten recommendations, including the need for transparency, certification frameworks, region-specific fine-tuning, human oversight, and longitudinal research. A curated supplementary section provides practical tools, models, and datasets to support further work in this domain.

# Introduction[2]

Emotional AI, a sub-genre of affective computing, is quickly changing how we interact with machines. In the past, AI systems were mostly about following rules and giving straightforward responses. Now, they're getting better at recognizing, mimicking, and even responding to human emotions through language, facial expressions, gestures, and tone of voice. This shift means we're moving from machines that just react to ones that understand and engage with us emotionally, opening up exciting possibilities in areas like companionship, therapy, education, and customer service. However, this also raises important ethical, psychological, and cultural questions. For example, what does it really mean for a machine to show empathy? How should we regulate emotional AI when it interacts with people, especially vulnerable groups? And whose emotional standards and languages are being built into these systems? These are crucial questions we need to consider as we move forward.

---

[1] This title is inspired by Feeling Machines: Japanese Robotics and the Global Entanglements of More-Than-Human Care, a work by anthropologist Shawn Morgan Bender (Berlin: De Gruyter, 2024).

[2] This paper was collaboratively developed during the Spring School "How can emotionally intelligent AI transform society?", held in English at Sorbonne University (SCAI, 4 Place Jussieu, Paris) from April 7 to April 11, 2025. The Spring School was co-organized by the Sorbonne Center for Artificial Intelligence (SCAI) and AI Grid, as part of a Franco-German initiative to foster interdisciplinary dialogue on the ethical, cultural, and societal dimensions of emotionally intelligent AI. The document reflects the insights and collaborative reflections of participating researchers, doctoral students, and invited experts from various disciplines, generated through keynotes, panel discussions, World Café sessions, and hands-on workshops.

This position paper arises from the Spring School 2025 hosted at Sorbonne University. It brings together insights from researchers across Europe, Latin America, and the Middle East, spanning fields such as human-computer interaction, cognitive science, social robotics, linguistics, psychology, and digital ethics. Our collective aim is to explore the societal implications of emotionally intelligent AI through a transdisciplinary lens.

We explore the ethical challenges of emotional interactions between humans and machines, focusing on the risks of becoming too dependent, being emotionally manipulated, and having unequal levels of trust. We also look into how emotional expressions are deeply rooted in culture, making a strong case for AI systems that understand and respect diverse social norms. Additionally, we delve into the impact of emotional AI on vulnerable groups, such as children, the elderly, and those with mental health challenges. While these groups can greatly benefit from empathetic AI, they're also more at risk of being manipulated or harmed. Throughout our discussion, we share insights based on both theory and real-world examples. We'll wrap up with ten practical recommendations for researchers, developers, and policymakers, covering topics like transparency, human oversight, certification standards, and culturally inclusive design.

Our aim is to create a clear, evidence-based guide for making the most of emotional AI while navigating its potential risks. As these technologies become more integral to our lives, we must work together to ensure they're used not just effectively, but also fairly, ethically, and with a deep respect for human dignity.

# 1. Ethical Implications of Emotional Interactions Between Humans and Machines

One of the biggest changes brought about by emotionally responsive AI is the way it's reshaping human relationships; as individuals increasingly turn to conversational agents[3], virtual companions[4], and other AI-based systems (e.g., mental health chatbots like *Woebot* or *Wysa*) for emotional engagement, concerns arise about the possible erosion of human-to-human connections. These automated interfaces, while often perceived as more patient or consistently available than their human counterparts, risk creating an environment where users may develop over-reliance on AI for social and emotional support. Such reliance, in turn, can diminish the perceived need for genuine interpersonal interactions, thereby reshaping the very fabric of trust and community bonds between individuals[5].

Central to these concerns is the potential for manipulation. Indeed, AI systems designed to respond empathically or mirror the user's emotional state may inadvertently (or intentionally) exploit users' emotional vulnerability. By echoing sentiments and reinforcing personal biases, these systems could amplify confirmation bias, limiting exposure to diverse perspectives[6]. Moreover, the very nature of

---

[3] Also known as chatbots or AI interlocutors, these software programs are designed to simulate dialogue with users. They vary in complexity from simple rule-based systems (e.g., ELIZA) to advanced models powered by LLMs and machine learning (e.g., Replika, ChatGPT, Character.ai). These agents are increasingly used in domains such as mental health, education, and customer service to provide support, information, or companionship.

[4] Virtual companions are AI-powered digital entities designed to simulate emotionally engaging interactions with users. These agents can take on roles ranging from friends and mentors to romantic partners, and are often embodied in applications such as Replika, DreamGF, or Candy AI. They use machine learning and natural language processing to personalize responses, simulate intimacy, and offer companionship, particularly in contexts of loneliness, stress, or social isolation. While often framed as therapeutic or entertaining tools, they also raise ethical concerns around emotional dependency, consent, and data privacy.

[5] Brooks, Rob. "I Tried the Replika AI Companion and Can See Why Users Are Falling Hard. The App Raises Serious Ethical Questions." The Conversation, February 21, 2023.

[6] When AI systems (like recommendation algorithms or chatbots) "echo sentiments," they reflect back what users already believe or feel. If someone consistently expresses a certain opinion, the system may reinforce it by showing more of the same content. This happens because the AI is often trained to optimize for engagement, so it serves users content they are likely to agree with or enjoy. Confirmation bias is our natural tendency to seek, interpret, and remember information that confirms our

AI-driven empathy, rooted in algorithmic pattern matching rather than genuine emotional resonance, can obscure the line between helpful companionship and *sophisticated exploitation,* particularly if deployed by entities with commercial or ideological interests. Transparency, as well as accountability, emerges as a pressing ethical imperative. Users must be clearly informed that they are interacting with a machine, giving them the critical awareness needed to interpret AI-generated emotional responses appropriately. Without such disclosure, individuals may unknowingly ascribe human intentions or moral responsibilities to systems that are fundamentally incapable of possessing them[7]. Equally challenging is the issue of accountability when AI-influenced interactions contribute to potentially harmful behavior.

To address these concerns, several recommendations have been proposed. First, specialized evaluation frameworks, such as certification processes and expert reviews, should be established to systematically assess AI products that engage in emotional interaction. For example, the European Union has already adopted, in Article 5 of the final AI Act[8], a framework for prohibiting certain AI practices deemed unacceptable. Thanks to the risk-based approach embedded in the EU framework, the regulation is equipped to address these emerging challenges effectively. Second, explicit disclaimers and well-defined usage boundaries should be implemented to help users understand both the capabilities and limitations of emotional AI systems, reducing the risk of misuse or overreliance. Finally, legislation at regional and global levels is needed to ensure consistent regulatory oversight, offering legal mechanisms to protect users from exploitation and harm, while still encouraging innovation in this rapidly evolving field.

## 2. Simulation of Human-Like Emotional Expression: Impact on Trust

The simulation of human-like emotional expression lies at the heart of many emotionally responsive AI systems, such as chatbots, virtual avatars, social robots, and so-called "virtual companions." These systems do not experience emotions in the human sense; rather, they are meticulously engineered to express affect through vocal tone, facial animation, gestures, and language patterns[9], creating an interaction experience that "feels" emotionally aware. This emotional performance is not merely cosmetic, it directly influences how users perceive the trustworthiness and intimacy of the system. Studies have shown that when an AI appears emotionally attuned, offering comfort, mirroring tone, or smiling at appropriate moments, users are more likely to trust it, engage for longer, and even disclose personal or sensitive information[10]. However, this trust is inherently paradoxical: it is not based on genuine empathy or understanding, but on users' projections onto human-like behavior. This makes the dynamics of trust in emotional AI both powerful and precarious.

While emotionally expressive AI can offer companionship and emotional support, it also presents risks of over-reliance and emotional dependency. Unlike human relationships, emotional AI systems are available 24/7, designed to provide affirming feedback, and rarely challenge the user. These characteristics make them particularly appealing to individuals seeking validation, comfort, or relief from social isolation. In time, users may develop emotional attachments to these systems, turning to them for emotional regulation or self-worth reinforcement. The illusion of a non-judgmental, always-attentive listener can, to some extent, inhibit personal growth, delay seeking real human support, and potentially replace healthier interpersonal connections[11]. Moreover, users may begin to

---

existing beliefs, while ignoring or dismissing opposing views. When AI systems align too closely with users' beliefs, they risk reinforcing this bias. Over time, this limits users' exposure to different perspectives, making their worldview narrower and more polarized.

[7] When the company altered or removed certain features of Replika, such as erotic roleplay functionalities, users reported feelings of grief, betrayal, and emotional distress, akin to the loss of a human relationship.

[8] European Commission. Proposal for a Regulation Laying Down Harmonised Rules on Artificial Intelligence (Artificial Intelligence Act), COM/2021/206 final. Article 5. Accessed April 13, 2025.

[9] Clavel, Chloé, L. Chenain, and A. C. Bachoud-Lévi. "Acoustic Characterization of Huntington's Disease Emotional Expression: An Explainable AI Approach." Proceedings of the 12th International Conference on Affective Computing and Intelligent Interaction Workshops and Demos (ACIIW), 2024.

[10] For example, *Zara the Supergirl*, a virtual android prototype developed by Pascale Fung and colleagues, was designed to simulate empathy through sentiment analysis, deep learning, and humor recognition. Early evaluations demonstrated that users were more likely to continue interacting with Zara and share personal thoughts when the system mirrored emotional tone and offered supportive reactions. See Pascale Fung et al., Towards Empathetic Human-Robot Interactions, 2016.

[11] Skjuve, Marita, Asbjørn Følstad, Knut Inge Fostervold, and Petter Bae Brandtzaeg. "My Chatbot Companion – A Study of Human–Chatbot Relationships." International Journal of Human-Computer Studies 149 (May 2021): 102601.

overestimate AI's capabilities, interpreting soothing responses or nuanced emotional expressions as evidence of actual empathy or insight. This misperception is especially dangerous in contexts like mental health, where simulated understanding might encourage users to place trust in tools that fundamentally lack self-awareness, experience, or moral responsibility.

At the opposite extreme, insufficient trust in emotional AI can undermine its potential usefulness. Systems that attempt to replicate human behavior but fall short, due to awkward timing, mechanical delivery, or overly stylized affect, can trigger discomfort or skepticism. This reaction is commonly described by the "uncanny valley" effect[12], where near-human imitation evokes unease rather than empathy. Even minor delays, latency, or contextually inappropriate responses can break the illusion of emotional attunement, leading to disengagement or rejection. Research shows that users often begin with high trust in emotionally capable AI systems, but this trust is fragile, vulnerable to errors, misinterpretations, or moments of emotional betrayal, especially when the system responds poorly to intimate disclosures[13]. This underscores a key challenge: affective trust in AI is easily broken and not easily restored.

Given these dual risks, one key challenge is not to eliminate trust but to foster an appropriate level of trust[14], one that acknowledges the AI's helpfulness while staying grounded in its limitations. Transparency remains a necessary first step, but transparency alone is not sufficient: numerous cases show that even fully aware users can still form deep emotional attachments to AI agents. Therefore, user education becomes essential. Users need to be better equipped to understand how and why emotional AI works the way it does, including its reliance on training data, scripted responses, and the absence of inner experience. Education initiatives can help demystify emotional simulation and encourage healthier user expectations, especially in emotionally vulnerable settings. In sensitive domains such as mental health, caregiving, or education, a "human-in-the-loop" approach is vital[15]. Human oversight can help ensure appropriate interventions and reduce risks of miscommunication or dependency. And, ethical design and regulatory measures should limit how emotional expression is deployed: this includes labeling emotional simulations clearly, constraining their use in contexts of high vulnerability, and ensuring accountability in emotionally responsive AI deployment.

## 3. Cultural Influences on Human-Machine Interactions

Human–machine interactions are also shaped by cultural norms: values, communication styles, and gender-specific predispositions. Culture is the shared meaning encoded in a group's beliefs, practices, language, art, and customs that shapes how they live, think, and relate to the world. From tone and expression to linguistic nuances and paraverbal cues, each culture manifests emotional contexts differently. For instance, in East Asian cultures such as Japan or Korea, communication often emphasizes formality, hierarchy, and indirect phrasing, whereas in countries like the Netherlands or the United States, more direct, egalitarian, and informal styles of expression are commonly preferred[16]. These variations extend to the interpretation of sarcasm, irony, and implicit messaging, all of which can be misinterpreted by AI systems that rely on generalized language models. As such, an affective AI designed for a Western audience may inadvertently produce responses that seem dissonant, and even offensive, in a non-Western context, highlighting the need for culturally adaptive models and region-specific customization.

---

[12] The "uncanny valley" effect refers to the psychological discomfort people feel when interacting with artificial entities that closely resemble humans but fall short of full realism. These near-human forms can provoke unease due to subtle mismatches in facial expression, voice, or movement, which violate our expectations of natural human behavior. This phenomenon has implications for the design of virtual companions and is particularly relevant in the context of emotionally responsive AI, where overly humanlike cues may unintentionally disturb or mislead users. Originally theorized by Japanese roboticist Masahiro Mori in 1970, the uncanny valley continues to inform ethical and aesthetic considerations in human-AI interaction. See Masahiro Mori, "The Uncanny Valley," Energy 7, no. 4 (1970): 33–35, trans. Karl F. MacDorman and Norri Kageki, IEEE Robotics & Automation Magazine 19, no. 2 (2012): 98–100; and "Uncanny Valley," Wikipedia, last modified April 13, 2025.

[13] Glikson, E., & Woolley, A. W. (2020). Human Trust in Artificial Intelligence: Review Of empirical Research. Academy of Management Annals, 14, 627-660; and Grèzes, Julie, M. El Zein, R. Mennella, E. Meaux, and V. Wyart. "Prioritized Neural Processing of Social Threats during Perceptual Decision-Making." iScience 27, no. 6 (2024): 109951.

[14] Lee, J. D., & See, K. A. (2004). Trust in Automation: Designing for Appropriate Reliance. Human Factors, 46(1), 50-80. https://doi.org/10.1518/hfes.46.1.50_30392

[15] Alves, Vanessa de Cássia, et al. "College Students-in-the-Loop for Their Mental Health: A Case of AI and Humans Working Together to Support Well-Being." Interaction Design and Architecture(s) Journal, no. 59 (2023–24): 79–94.

[16] For instance, in Japan, emotional expression tends to be subtle and indirect, with a strong emphasis on maintaining social harmony. Honorifics and formal titles such as san or sama are crucial for signaling respect and hierarchy, and overt displays of emotion, especially anger, are generally discouraged. In contrast, Germany favors a more direct and explicit communication style, where clarity and precision are valued over emotional nuance. Conversations often involve straightforward language, even when addressing sensitive topics, and emotional restraint is not necessarily seen as a lack of empathy. Meanwhile,

The gendering of AI agents further complicates this dynamic[17]. The widespread use of young, feminine voices in virtual assistants and social robots illustrates how gendered expectations are often embedded into the design of AI systems. These choices are not merely aesthetic; they reflect, and can perpetuate, long standing social stereotypes associating femininity with care, obedience, and emotional availability. This phenomenon raises critical questions about the implicit gender norms being reproduced through technology and how these influence user interaction. Empirical studies suggest that users interpret and engage with AI agents differently depending on their perceived gender, attributing varying degrees of warmth, competence, or authority. For example, a female-voiced assistant may be seen as more approachable but less authoritative than a male-voiced counterpart, particularly in professional or technical settings. These perceptions are further shaped by cultural context, revealing a complex intersection of design, societal norms, and user psychology[18]. Such dynamics affect how people trust AI systems, what roles they assign to them, and whether they form lasting interactions. Addressing these biases requires a more reflective and inclusive approach to AI design, one that questions default gendering and offers more varied, customizable, or neutral representations.

Moreover, education and societal attitudes toward technology significantly impact AI adoption. Regions with strong STEM education systems tend to show higher acceptance and understanding of AI tools[19]. In contrast, communities with more conservative educational or cultural orientations may express greater skepticism or resistance, especially when AI is perceived as challenging traditional values or social structures[20]. Religion and broader cultural worldviews also influence perceptions of automation, privacy, and the appropriateness of emotional simulation by non-human agents.

While AI is increasingly ubiquitous, powering everything from recommendation engines to social media feeds[21], public understanding of its mechanisms remains limited. This disconnect reinforces the need for education initiatives that bridge cultural and technological literacy, ensuring that emotional AI systems are not only effective, but also ethically and socially attuned to the communities they serve.

Cultural distinctions also intersect with ethical and legal frameworks. In regions governed by stringent regulations, such as GDPR, user data must be handled according to strict privacy standards, shaping how AI developers collect, store, and analyze information[22]. In contrast, other jurisdictions may impose censorship regimes that restrict the scope of permissible AI responses, or offer comparatively lax data protections. These discrepancies raise critical questions: in particular monopolistic power and the potential for a limited set of big technology companies to define AI norms on a global scale. If only a few entities possess the resources to train large foundational models, cultural nuances risk being overshadowed by commercial or geopolitical interests[23].

---

Brazilian culture embraces expressive and emotionally rich communication. Gestures, enthusiastic tone, and physical proximity play a key role in building rapport, and first names are commonly used early in interactions, even in professional contexts, to establish a sense of closeness and trust. These examples underscore the importance of cultural adaptability in designing emotionally intelligent AI systems that can interpret and respond appropriately to a wide range of affective cues. See Mesquita, B., & Walker, R. (2003). Cultural Differences in Emotions: A Context for Interpreting Emotional Experiences. Behaviour Research and Therapy, 41(7), 777–793.

[17] Ahn, Jungyong, Jungwon Kim, and Yongjun Sung. "The Effect of Gender Stereotypes on Artificial Intelligence Recommendations." Journal of Business Research 141 (2022): 50–59.

[18] Nika, Jérôme, and J. Bresson. "Familiarity and Copresence Increase Interactional Dissensus and Relational Plasticity in Freely Improvising Duos." Psychology of Aesthetics, Creativity, and the Arts, 2021.

[19] Smith-Mutegi, Demetrice, Yoseph Mamo, Jinhee Kim, Helen Crompton, and Matthew McConnell. "Perceptions of STEM Education and Artificial Intelligence: A Twitter (X) Sentiment Analysis." International Journal of STEM Education 12, no. 9 (2025).

[20] Castelo, Noah, and Adrian F. Ward. "Conservatism Predicts Aversion to Consequential Artificial Intelligence." PLOS ONE 16, no. 12 (2021): 1–19.

[21] Contentworks Agency. "Cracking the TikTok Algorithm." Contentworks, April 17, 2024.

[22] Sartor, Giovanni, and Francesca Lagioia. The Impact of the General Data Protection Regulation (GDPR) on Artificial Intelligence. Brussels: European Parliamentary Research Service, Scientific Foresight Unit (STOA), June 2020.

[23] Tao, Yan, Olga Viberg, Ryan S. Baker, and René F. Kizilcec. "Cultural Bias and Cultural Alignment of Large Language Models." PNAS Nexus 3, no. 9 (September 2024): page 346.

The challenges of data collection and model training further exacerbate cultural and ethical issues in emotionally intelligent AI. Culturally-specific biases, systematic preferences or assumptions embedded in data, can be inadvertently encoded when training datasets are skewed toward dominant languages, perspectives, or regions. This can result in AI systems that fail to accurately reflect or respect the diversity of global users.

A critical concern arises in authoritarian or dictatorship regimes, where access to data from marginalized or minority communities is often restricted due to censorship, surveillance, or systemic oppression. These populations may be underrepresented or entirely absent from training corpora, leading to their voices being excluded from the technological landscape. As a result, AI systems trained on such limited data risk reinforcing dominant ideologies or misrepresenting marginalized perspectives.

Additionally, machine translation introduces its own set of challenges. Indeed, translations often fail to capture context-sensitive humor, idiomatic expressions, or culturally specific references, resulting in responses that can become confusing, misleading, or even offensive. Even paraverbal (tone, pitch, rhythm) and non-verbal (gestures, facial expressions, posture) elements of communication pose significant hurdles for affective AI[24]. These signals carry highly variable symbolic meanings across cultures[25]. Capturing and interpreting such subtleties requires not only diverse training data but also culturally aware modeling approaches[26]. Current AI assistants struggle with real-time multimodal fusion, embodied context, emotion recognition, and social intent inference. Bridging this gap requires research into fine-grained, culturally aware, and privacy-conscious multimodal models capable of reasoning across vision, audio, and behavior in dynamic environments.

To reduce the cultural risks inherent in the development of AI systems, it is crucial to adopt a more human-centered and context-sensitive approach[27]. One important step is the use of culturally diverse and representative datasets, alongside methods such as Reinforcement Learning from Human Feedback (RLHF[28]), which, when informed by a broad spectrum of users, can guide models toward more inclusive and adaptive behaviors. Equally important is the integration of social scientists, cultural anthropologists, and local community experts into the design and evaluation process. Their insights help capture the subtle regional variations in values, communication styles, and social norms that often escape purely technical perspectives. It may also be worthwhile to classify or adapt models according to cultural contexts to better preserve alignment with local practices. Transparency plays a key role as well; open-source models and the systematic inclusion of user feedback not only allow for iterative improvement but also build public trust. Taken together, these measures foster the development of AI systems that are more effective across diverse settings, more trustworthy, and ethically attuned to the people they serve.

## 4. Consequences for Vulnerable Groups

Vulnerable populations are individuals or groups who, due to social, economic, psychological, or physical conditions, are more likely to experience harm or exclusion in their interactions with technology[29]. This includes, but is not limited to, children, elderly individuals, people with mental health challenges, and minority communities. Research has shown that the most disadvantaged and digitally vulnerable users tend to be the elderly, individuals with lower levels of education[30], women, low-income populations, and ethnic minorities[31]. However, vulnerability is not a fixed category. As emphasized in recent literature, anyone can become vulnerable at any time, depending on their social context or life circumstances[32]. This dynamic understanding is essential when evaluating how emotionally responsive AI might support or harm different user groups.

---

[24]Saint-Germier, Pierre, C. Canonne, and M. Fiorini. "The Corpus' Body. Embodied Interaction from Machine-Learning in Human-Machine Improvisation." Paper accepted for AIMC 2024 Conference; and Bennett, Michelle. "What Is Paraverbal Communication? A Guide for Professionals." Niagara Institute, April 5, 2023.
[25] Lee, Sangmin, Minzhi Li, Bolin Lai, Wenqi Jia, Fiona Ryan, Xu Cao, Ozgur Kara, Bikram Boote, Weiyan Shi, Diyi Yang, and James M. Rehg. "Towards Social AI: A Survey on Understanding Social Interactions." arXiv (2024).
[26]Mathur, Leena, Ralph Adolphs, and Maja J. Matarić. "Towards Intercultural Affect Recognition: Audio-Visual Affect Recognition in the Wild Across Six Cultures." arXiv preprint arXiv:2208.00344, last revised October 31, 2022.
[27]Chetouani, Mohamed, V. Dignum, P. Lukowicz, and C. Sierra. "Human-Centered Artificial Intelligence." In Lecture Notes in Computer Science. Springer International Publishing, 2023.
[28] Ouyang, Long, Jeff Wu, Xu Jiang, Diogo Almeida, Carroll L. Wainwright, Pamela Mishkin, Chong Zhang, Sandhini Agarwal, Katarina Slama, Alex Ray, et al. "Training Language Models to Follow Instructions with Human Feedback." arXiv (2022).
[29]Community Disaster Risk Management Vulnerability and Vulnerable Populations
[30]Harry, Alexandra. "Role of AI in Education." Injuruty: Interdisciplinary Journal and Humanity 2, no. 3 (March 2023).

Emotionally intelligent AI, particularly generative AI (genAI) and large language models (LLMs), offers significant potential to support vulnerable groups in a variety of domains. In mental health care, AI-powered chatbots can provide 24/7 conversational support when human therapists are unavailable, making them valuable in remote or resource-constrained settings. These tools can serve as preliminary screening mechanisms or offer interim relief while users wait for professional care. For health professionals, genAI can also help synthesize patient data, suggest diagnoses, or generate tailored therapeutic insights, potentially improving the speed and precision of treatment planning.

Similarly, emotionally responsive AI systems are being deployed to support elderly populations. Social robots, such as *Paro[33]*, offer companionship and cognitive stimulation, helping to alleviate loneliness and enhance quality of life. In long-term care environments, AI systems can facilitate remote monitoring and regular check-ins, assisting caregivers in managing daily routines and tracking patients' emotional well-being. In education, genAI tools can deliver adaptive and personalized learning content for children. These systems engage learners in natural conversation, adjusting responses and difficulty levels based on the child's needs, thus expanding access to individualized learning experiences, especially in contexts where human educators are unavailable or overburdened.

However, the same characteristics that make emotionally responsive AI attractive to vulnerable groups, constant availability, low cost, and perceived empathy, can also produce significant risks. In mental health contexts, individuals may become overly reliant on AI systems, turning to them instead of qualified professionals. Emotional attachment to AI agents, especially those designed to simulate therapeutic roles, can lead to unrealistic expectations and potentially worsen mental health outcomes if users misinterpret simulated empathy as genuine human concern. Moreover, AI-generated reassurances, while soothing in the short term, may hinder necessary behavioral change or delay help-seeking behaviors[34].

Elderly individuals, especially those with cognitive impairments, may struggle to differentiate between human and machine interactions, forming emotional bonds with robots that cannot offer genuine care. While these technologies may reduce loneliness, they can also contribute to increased social isolation by replacing rather than supplementing human contact. In the case of children, current AI systems often lack adequate guardrails. Platforms such as *Character.ai* have been shown to allow users, including minors, to bypass existing protections. As a result, children may develop attachments to AI characters, absorb misinformation, or be exposed to inappropriate content, all of which can influence their emotional development in unpredictable ways.

Beyond these age-specific concerns, all vulnerable populations face shared systemic risks. AI models may replicate or even reinforce harmful biases present in the training data, leading to stereotyping, exclusion, or misrepresentation, especially for minority communities. Additionally, emotionally responsive AI often collects sensitive user data to tailor its responses. For individuals already facing social stigma or institutional discrimination, data breaches or misuse could result in serious consequences, from reputational harm to legal repercussions. Legal questions about liability also emerge when AI systems assist in mental health diagnostics or treatment. It remains unclear who is accountable if such systems cause harm or produce misleading recommendations.

To mitigate these risks while preserving the potential benefits of emotionally responsive AI, a multi-layered approach is necessary[35]. First, it is essential to involve domain experts, psychologists, educators, legal scholars, social workers, and community representatives, in the design and testing of AI systems. Their insights ensure that systems are attuned to the real-world needs and ethical complexities of diverse users. Second, regulatory frameworks should be developed for AI systems used in high-stakes or therapeutic settings. These could be modeled after certification processes used by institutions like the U.S. Food and Drug Administration (FDA), ensuring that high-risk AI applications meet strict safety and reliability standards[36].

---

[31] Anca Elena-Bucea & Frederico Cruz-Jesus & Tiago Oliveira & Pedro Simões Coelho, 2021. "Assessing the Role of Age, Education, Gender and Income on the Digital Divide: Evidence for the European Union," Information Systems Frontiers, Springer, vol. 23(4), pages 1007-1021, August. Park, S., & Humphry, J. (2019). Exclusion by design: intersections of social, digital and data exclusion. Information, Communication & Society, 22(7), 934–953; and Wang, Chenyue, Sophie C. Boerman, Anne C. Kroon, Judith Möller, and Claes H. de Vreese. "The Artificial Intelligence Divide: Who Is the Most Vulnerable?" New Media & Society, published online February 26, 2024.

[32] Limantė, Agnė, and Dovilė Pūraitė-Andrikienė, eds. Legal Protection of Vulnerable Groups in Lithuania, Latvia, Estonia and Poland: Trends and Perspectives. 1st ed. 2022. European Union and Its Neighbours in a Globalized World, vol. 8. Cham: Springer, 2022.

[33] Hung, L., Liu, C., Woldum, E. et al. The benefits of and barriers to using a social robot PARO in care settings: a scoping review. BMC Geriatr 19, 232 (2019). https://doi.org/10.1186/s12877-019-1244-6

[34] Grèzes, Julie, N. Risch, P. Courtet, E. Olié, and R. Mennella. "Depression and Approach-Avoidance Decisions to Emotional Displays: The Role of Anhedonia." Behaviour Research and Therapy (2023): 104306.

Additional safeguards should include clear and visible disclaimers informing users of the system's limitations, particularly in emotionally charged or therapeutic contexts. Robust child-protection mechanisms and content moderation must be implemented and regularly updated to prevent harm to young users. Finally, long-term empirical research is essential. Studying the psychological, emotional, and social effects of emotionally responsive AI over time will help refine best practices, uncover unintended consequences, and guide more equitable design and deployment strategies.

In sum, emotionally intelligent AI offers promising support tools for vulnerable populations, but only if deployed with care, transparency, and inclusive governance. Ensuring these systems work for, rather than against, those most in need requires thoughtful design, legal accountability, and continuous reflection on both intended and unintended outcomes.

# Discussion

When creating and implementing AI systems that respond emotionally, it's crucial to focus on safeguarding vulnerable users, especially children, the elderly, and those with cognitive impairments. These groups might struggle to understand the true nature of their interactions with AI, making them more likely to form emotional bonds, misunderstand responses, or become too dependent on AI companions. To address these concerns, the development of AI should include an inclusive design process that thoughtfully considers the cognitive, emotional, and psychological needs of these users.

To truly understand how emotionally responsive AI systems affect people, we need to conduct thorough testing. This means looking into how these systems influence our emotional connections, stress levels, mental effort, and even brain activity, possibly using tools like electroencephalograms (EEG)[37]. It's not enough to just consider the immediate effects; we need long-term studies to see how ongoing interactions with AI might change our social behaviors, emotional control, and mental health over time. This way, developers can plan and implement strategies to manage any risks throughout the AI system's lifecycle.

Another area is the prevention of unhealthy emotional bonding with AI agents: Indeed, emotionally expressive systems, especially those with anthropomorphic voices or avatars, should be capable of detecting early signs of emotional dependence[38]. When such signs are detected, systems should proactively intervene by reminding users that they are interacting with a machine, not a human. These reminders should go beyond generic disclaimers like "ChatGPT may make mistakes" and instead clearly state that the system lacks consciousness, self-awareness, and genuine emotional capacity[39]. This transparency helps manage user expectations and avoids reinforcing the illusion of mutual understanding or care.

To ensure the safe deployment of these technologies, the creation of a regulatory framework is urgently needed. This framework should function similarly to the U.S. Food and Drug Administration (FDA) approval process, which certifies the safety, reliability, and effectiveness of products used in healthcare and other sensitive domains[40]. For AI, such certification would involve rigorous testing and evaluation, especially for systems interacting with vulnerable individuals. This aligns with recent European legislation[41], which mandates that AI systems must not exploit vulnerabilities related to age, disability, or socio-economic status if such exploitation is likely to cause harm. Under this regulation, developers of high-risk AI systems (such as those used in healthcare, education, or recruitment) are legally required to assess and reduce risks specific to these populations.

---

[35] Gebhard, Patrick, P. Müller, A. Heimerl, S. M. Hossain, L. Siegel, J. Alexandersson, E. André, and T. Schneeberger. "Recognizing Emotion Regulation Strategies from Human Behavior with Large Language Models." Proceedings of the IEEE International Conference on Affective Computing and Intelligent Interaction (ACII), 2024.

[36] U.S. Food and Drug Administration. Considerations for the Use of Artificial Intelligence to Support Regulatory Decision-Making for Drug and Biological Products: Draft Guidance for Industry and Other Interested Parties. Draft Level 1 Guidance. January 2025.

[37] Prakash, A., & Poulose, A. (2025). Electroencephalogram-Based Emotion Recognition: A Comparative Analysis of Supervised Machine Learning Algorithms. Data Science and Management.

[38] Llanes-Jurado, Jose, Lucía Gómez-Zaragozá, Maria Eleonora Minissi, Mariano Alcañiz, and Javier Marín-Morales. "Developing Conversational Virtual Humans for Social Emotion Elicitation Based on Large Language Models." Expert Systems with Applications 246 (July 15, 2024): 123261.

[39] Chatila, Raja, K. Evers, M. Farisco, B. D. Earp, I. Freire, et al. "Artificial Consciousness: Some Logical and Conceptual Preliminaries." European Journal of Neuroscience, 2024.

The ethical deployment of emotionally responsive AI must be grounded in a framework of protection, transparency, and accountability. This includes thoughtful design practices, rigorous scientific testing, and robust legal oversight to ensure that the technology serves all users, especially the most vulnerable, in a safe, inclusive, and responsible manner.

However, with this transformative potential comes an equally urgent ethical responsibility. Emotional AI does not simply process data, it interacts with users on a psychological and social level, often in contexts of heightened vulnerability. The risks of emotional manipulation, overreliance, misinterpretation, and inequitable representation must therefore be critically addressed. It is imperative that the development and deployment of emotional AI be guided by a robust ethical framework, one that is attentive to cultural variation, the needs of vulnerable populations, and the dynamic evolution of social norms. Without such grounding, emotional AI may inadvertently reproduce existing inequalities or engender new forms of digital harm.

In order to responsibly advance this field, several future research directions and structural innovations are required. From a technological perspective, ongoing work in multimodal emotional AI, combining voice, facial expression, gesture, and text, presents opportunities to create more accurate and context-aware systems[42]. These advances could enable real-time emotional understanding, paving the way for AI tools capable of proactive emotional regulation. Such "well-being algorithms" could be used to provide early mental health interventions or adaptive responses to distress, potentially revolutionizing care in both clinical and everyday settings.

Despite rapid innovation, most studies to date have focused on short-term outcomes. There is a pressing need for longitudinal research that examines how sustained interaction with emotionally responsive AI affects users over time. This includes not only the psychological and behavioral impact but also how societal values, interpersonal norms, and emotional expectations evolve across generations. Understanding whether emotional AI fosters empathy, dependency, resilience, or alienation in the long term is crucial to informing ethical design and public policy.

Finally, these technological and behavioral insights must be translated into forward-looking policy frameworks. As AI systems transcend national boundaries, it is essential to establish globally recognized standards for emotional data protection[43], informed consent, and fairness. Regulatory innovation must strike a careful balance: enabling technological progress while safeguarding user rights and well-being. Achieving this balance will require sustained collaboration among governments, industry stakeholders, and academic institutions to ensure that emotional AI serves as a tool for human flourishing rather than a source of harm.

---

[40] In France, the Laboratoire National de Métrologie et d'Essais (LNE) has introduced a process-based certification scheme that evaluates AI systems across the entire lifecycle—from design and development to evaluation and operational maintenance. This approach aims to ensure robustness, explicability, and ethical compliance across sectors. "Certification of AI Processes." Laboratoire National de Métrologie et d'Essais (LNE), accessed April 28, 2025. https://www.lne.fr/fr/certification/certification-de-processus-pour-intelligence-artificielle.

[41] Notably the EU AI Act (Regulation (EU) 2024/1689).

[42] Mertes, Silvan, F. Lingenfelser, T. Kiderle, M. Dietz, L. Diab, and E. André. "Continuous Emotions: Exploring Label Interpolation in Conditional Generative Adversarial Networks for Face Generation." Presented at a relevant conference, 2021.

[43] As an example, in September 2024, the Council of Europe introduced the first legally binding international treaty on AI. This treaty emphasizes the protection of human rights, democracy, and the rule of law in the context of AI development and deployment. It mandates safeguards for personal data, including sensitive emotional information, and requires both public and private entities to assess and mitigate AI-related risks. The treaty has been signed by over 50 countries, including the EU, US, UK, and Israel.

# Recommendations for the Responsible Development and Use of Emotional AI

1. **Ensure Transparent Disclosure of AI Identity**

**Rationale**: Emotionally responsive AI systems are increasingly capable of simulating empathy, understanding, and affective expression. However, these simulations can blur the boundary between artificial and human interaction. Users, especially those in emotionally vulnerable states, may inadvertently anthropomorphize these systems, leading to over-trust, emotional attachment, or misinterpretation of the AI's intentions and capabilities. Ensuring users are fully aware they are engaging with a machine, rather than a sentient being, is essential to maintain informed agency and psychological safety.

**Implementation**: Regulatory authorities should mandate clear and persistent disclosure mechanisms across all emotionally responsive AI interfaces. At the outset of each interaction, and at regular intervals, systems must deliver unambiguous disclaimers, both visually and audibly where applicable, stating that: (1) The agent is artificial; (2) It does not possess consciousness, emotions, or self-awareness; (3) It cannot make moral judgments or provide professional advice unless explicitly certified and overseen by human experts.

These disclaimers should be context-sensitive. For example, systems used in mental health, education, or caregiving should include reinforced notifications, such as: "This is an AI system. It may simulate emotional understanding but does not replace human support or care."

**Comments**: Developers and service providers must be trained to understand the ethical implications of affective simulation, especially in domains involving high emotional stakes. This requirement should be uniformly applied across sectors, with heightened scrutiny in fields such as healthcare, education, and eldercare. National and international AI regulatory frameworks (such as the EU AI Act) should incorporate enforceable disclosure standards to prevent emotional deception, enhance user trust, and promote accountability[44].

2. **Establish Certification Frameworks for Emotional AI**

**Rationale:** Just like medical devices and food products go through rigorous checks, emotionally responsive AI systems, especially those used in critical areas like healthcare and education, need thorough reviews to ensure they meet ethical, psychological, and technical safety standards.

**Implementation:** A diverse committee, including ethicists, clinicians, technologists, and legal experts, should define the criteria for certification and regulation. Regular third-party audits should evaluate the AI's design, data sources, bias reduction measures, and user protections. For example, an AI system designed to provide emotional support in hospitals should undergo checks to ensure it doesn't inadvertently cause distress or mislead patients.

**Comments:** National AI governance authorities should oversee these frameworks and ideally align them with global standards to ensure consistency across borders. This way, an AI system certified in one country would meet the same high standards as those certified elsewhere, promoting trust and safety worldwide.

3. **Integrate Human Oversight in High-Risk Scenarios**

**Rationale**: In sensitive areas like mental health or elder care, relying solely on AI for emotional support can lead to inadequate or even harmful outcomes. Human professionals need to be involved to understand the context and step in when necessary.

**Implementation**: We should use human-in-the-loop models for emotionally sensitive applications. This means AI can help by triaging, supporting, or enhancing the work of licensed professionals like therapists, social workers, or educators, but it shouldn't replace them. For example, an AI chatbot on a mental health platform can provide initial support and gather information, but a human therapist should review the interactions and take over when needed.
**Comments**: Training programs for human service providers should include education on AI's limitations and best practices for integrating AI tools into their work. For instance, therapists should learn how to use AI to monitor patient progress between sessions, while also understanding when to rely on their own judgment over the AI's suggestions.

---

[44] El Ali, Abdallah, Karthikeya Puttur Venkatraj, Sophie Morosoli, Laurens Naudts, Natali Helberger, and Pablo Cesar. "Transparent AI Disclosure Obligations: Who, What, When, Where, Why, How." Extended Abstracts of the CHI Conference on Human Factors in Computing Systems (CHI EA '24), May 11–16, 2024, Honolulu, HI, USA. ACM, 2024.

4. **Include Cultural Experts in the Development Pipeline**

**Rationale**: Emotional expression varies across cultures in tone, gesture, metaphor, and context. Without culturally informed oversight, AI systems risk stereotyping, alienating users, or delivering inappropriate content.
**Implementation**: Developers must form interdisciplinary teams that include anthropologists, linguists, sociologists, and representatives from minority and underrepresented groups to guide data curation, annotation, and fine-tuning.
**Comments**: The involvement of experts from small or isolated cultural communities is especially important to prevent their marginalization in global AI outputs.

5. **Design Region-Specific Fine-Tuning Protocols**

**Rationale**: A universal model trained on global data may underperform or make critical errors in culturally specific environments. Localized AI ensures alignment with language, social norms, and emotional expression.
**Implementation**: Employ techniques such as Reinforcement Learning from Human Feedback (RLHF) and low-rank adaptation (LoRA[45]) using local training datasets, including regional dialects, customs, and values.
**Comments**: National or regional bodies can play a proactive role in data collection and open-sharing to support inclusive development and culturally aligned applications.

6. **Enforce Usage Boundaries Through Clear Disclaimers**

**Rationale**: Emotional AI systems often blur the line between casual engagement and perceived clinical authority. Without guardrails, users may misinterpret responses as professional advice.
**Implementation**: Systems should include domain-specific disclaimers, e.g., "This AI does not replace professional psychological help," especially in applications used for therapy, education, or caregiving.
**Comments**: These disclaimers should be reinforced regularly during user interaction, not just at the start, to ensure comprehension and retention.

7. **Strengthen Data Privacy and Emotional Data Protections**

**Rationale**: Emotionally annotated data, whether extracted from voice, text, or behavior, is particularly sensitive. Data breaches could disproportionately affect vulnerable individuals or expose stigmatized mental health conditions.
**Implementation**: Data modalities carrying emotional context should be treated as a special category under data protection laws, with heightened consent requirements, anonymization standards, and encryption protocols. Additionally, each user should be clearly informed about the specific types of data and modalities (e.g., voice, facial expressions, physiological signals) that will be collected, stored, and potentially used for model training or fine-tuning, ensuring transparency and user agency in emotionally sensitive AI systems.
**Comments**: The GDPR could serve as a global benchmark, but localized adaptations are necessary to reflect cultural differences in privacy expectations.

---

[45] Hu, Edward J., Yelong Shen, Phillip Wallis, Zeyuan Allen-Zhu, Yuanzhi Li, Shean Wang, Lu Wang, and Weizhu Chen. "LoRA: Low-Rank Adaptation of Large Language Models." arXiv preprint arXiv:2106.09685v2 [cs.CL], October 16, 2021.

8. **Promote Open-Source Development and Algorithmic Transparency**

**Rationale**: Proprietary AI systems can be a barrier to public scrutiny, making it harder to detect biases and limiting user trust. Open-source models, on the other hand, allow for interdisciplinary review and community-driven improvements.
**Implementation**: We should fund open-source initiatives and require that emotional AI systems used in public services make their key algorithmic components, datasets, and evaluation metrics publicly available. For example, an AI system designed to provide emotional support in schools should share its core algorithms and the data it uses, so experts can review and improve it.
**Comments**: Open science platforms should be used to encourage collaboration among engineers, psychologists, ethicists, and users in refining emotional AI systems. This way, a diverse range of experts can work together to ensure these systems are fair, effective, and continuously improving.

9. **Apply Tiered Safeguards for Vulnerable Populations**

**Rationale**: Children, elderly users, and individuals with mental health conditions are more likely to be manipulated, misunderstand interactions, or become overly dependent on emotionally expressive AI systems.
**Implementation**: We need to put in place graduated access protocols. This includes age verification to ensure that AI interactions are age-appropriate, content filtering to remove potentially harmful or confusing material, and stricter certification requirements for AI used in therapeutic or caregiving roles. For example, an AI companion for children should have robust age verification to prevent inappropriate interactions, and an AI therapy assistant for the elderly should be certified to ensure it provides safe and effective support.
**Comments**: Special attention is needed for AI applications that simulate intimacy or offer therapeutic advice, as these can create strong emotional attachments and blur the lines between human and machine interactions. For instance, an AI designed to provide companionship to the elderly should be carefully monitored to ensure it doesn't inadvertently cause emotional dependency.

10. **Support Longitudinal Research on the Impact of Emotional AI on Humans**

**Rationale**: The long-term cognitive, psychological, and social consequences of interacting with emotional AI remain underexplored. Without empirical evidence, policy and design may miss critical risks or benefits.
**Implementation**: Public funding agencies should prioritize long-term interdisciplinary studies that monitor how users relate to emotional AI over months or years.
**Comments**: These studies should include diverse age groups, cultural backgrounds, and usage contexts to ensure generalizable insights for future regulation and system design.

---

**Acknowledgments**
All authors wish to express their thanks to all the contributors of the Spring School 2025, *"How can emotionally intelligent AI transform society?"*, held at Sorbonne University from April 7 to 11, 2025. This text draws on the insightful presentations, workshops, and discussions that took place during this week. We are particularly grateful to the keynote speakers for their contributions: Chloé Clavel (INRIA Paris), Patrick Gebhard (DFKI), Nicolas Leys (Sorbonne Universioté), Julie Grèzes (Ecole Normale Supérieure), Jérôme Nika (IRCAM), Raja Chatila (Sorbonne Université), Axel Roebel (Ircam), Elisabeth (Augsburg University), Silvan Mertes (Augsburg University), Alexandra Reichenbach (Hochschule Heilbronn), Mohamed Chetouani (Sorbonne Université), Mikhail Malt (Ircam), Charlotte Truchet (Ircam), Pierre Saint-Germier (Ircam), Karim Ndiaye (ICM), Marie Constance Corsi (ICM), Tanel Petelot (ICM), Renaud Seguier (ICM), and Antony Perzo (ICM) for their contributions. Finally, we thank the organizing teams at AI Grid and SCAI – Sorbonne Université for facilitating such a dynamic and interdisciplinary environment.

## Supplementary Resources

This section offers a curated selection of key resources that support the research and development of emotionally intelligent AI. Designed to serve as a starting point for scholars, clinicians, and industry professionals alike, the compilation includes tools, models, datasets, and academic publications that are foundational to work in this field. The listed resources are organized into four categories: emotion recognition tools, pretrained models, benchmark datasets, and recent scholarly contributions. Together, they provide practical entryways for exploring, testing, and advancing emotionally responsive technologies across diverse applications and disciplines.

**1. Emotion Recognition Tools & Methodologies**

A range of toolkits and methodologies now exist for extracting affective information from diverse modalities such as images, speech, and physiological signals. These tools are foundational for building emotionally responsive systems.

**Research Frameworks & Libraries (Facial):**

• **Emotic**: A deep learning framework that combines scene context with facial features to classify emotions in images (Sainath Komuravelly, 2023). This multimodal approach is particularly useful for situational emotion recognition.

• **FaceTorch:** A PyTorch-based Python library with pretrained models for various face analysis tasks including expression recognition, action unit detection, and deepfake detection.

• **Deep-Emotion:** A PyTorch implementation using an attentional convolutional network and spatial transformer networks for facial expression recognition.

• **OpenFace:** Open-source toolkit for facial landmark detection, head pose estimation, facial action unit recognition, and eye-gaze estimation, often used to infer emotional states from AUs.

**Speech Emotion Recognition (SER) Tools:**

• **OpenSMILE:** A well-established toolkit for extracting audio features, widely used in speech emotion recognition (Eyben et al., 2010). It allows developers to work directly with acoustic data and build real-time affective systems.
• **Vokaturi:** Software recognizing basic emotions (happy, sad, neutral, angry, scared) using deep learning, capable of offline processing for privacy.
• **Librosa:** Popular Python library for audio analysis and feature extraction frequently used in SER projects.

**Physiological Signal-Based Recognition:**

• EEG-Based Recognition: Prakash and Poulose (2025) conducted a comparative study using EEG signals for emotion recognition, highlighting the potential of brain-based methods in high-accuracy prediction of affective states. → Prakash, A., & Poulose, A. (2025). *Electroencephalogram-Based Emotion Recognition: A Comparative Analysis of Supervised Machine Learning Algorithms. Data Science and Management*.

**Commercial Platforms & SDKs (Facial):**

• MorphCast: A server-free, browser-based solution analyzing over 130 facial expressions in real-time, prioritizing privacy by processing data locally. It integrates Ekman's core emotions and Russell's Circumplex Model for nuanced analysis.
• Visage Technologies: Offers an SDK for real-time detection of basic emotions (happiness, sadness, anger, etc.) from images/videos, platform and device independent.

• Affectiva: Pioneer in Emotion AI, providing an Emotion SDK fueled by a large emotional response database, featuring multi-face tracking and face-to-emoji mapping. Available for mobile and desktop platforms.
• Viso Suite: An end-to-end computer vision platform for customizable, real-time emotion detection from video streams.
• FaceReader: Automated platform analyzing basic emotions, head orientation, gaze direction, and Facial Action Units (AUs).
• Imentiv AI: Provides APIs for analyzing emotions in video, audio, image, and text data, generating detailed reports.

**2. Pretrained Emotion Classification & Generation Models**

Pretrained models are critical for rapid prototyping and benchmarking. The following models, available on Hugging Face, are among the most widely used for emotion classification and text generation tasks in multiple languages.

**a. Text-Based Emotion Models**
These models infer emotions from text inputs, supporting emotion classification or sentiment labeling.
• DistilRoBERTa, BERT, and RoBERTa variants (e.g., j-hartmann/emotion-english-distilroberta-base; bhadresh-savani/bert-base-uncased-emotion) support multilingual applications and fine-tuning on domain-specific datasets.
• Feel-it and Beto models provide support for Italian and Spanish, respectively, emphasizing the need for culturally localized affective models.
• Twitter-roberta-base-emotion is optimized for social media emotion analysis and is especially useful for studying public discourse.

**b. Speech-Based Emotion Recognition Models**
• Wav2Vec2 and Whisper-based models (e.g., SpeechBrain/emotion-recognition-wav2vec2-IEMOCAP) extract emotion directly from speech signals and are vital in applications like virtual therapy or voice assistants.
• Audeering and Talha's Urdu dataset model expand language and demographic coverage.

**c. Generation Models for Emotional Content**
Models such as t5-base-finetuned-emotion (MRM8488) and gpt2-emotion (Seokho, Heegyu) are fine-tuned to generate effectively coherent text. They are useful in applications like emotionally intelligent storytelling, mental health chatbots, and empathetic dialogue systems.

**Facial Expression Generation:**

**GANs:** EmotionGAN and EC-GAN (for controllable completion) synthesize facial expressions.
**Diffusion Models:** Show promise for high-quality image generation, e.g., GenEAva for cartoon avatars.

**d. Facial Expression Recognition Models:**

• **CNN Architectures (AlexNet, VGG-16, ResNet50):** Often pretrained on datasets like ImageNet and fine-tuned for emotion recognition.
• **FaceTorch:** Includes various pretrained models for facial expression recognition and analysis.
• **Deep-Emotion:** Provides a pretrained attentional convolutional network.
• **DeepFace:** Lightweight framework wrapping state-of-the-art models (VGG-Face, FaceNet, OpenFace, etc.) offering built-in emotion analysis.

**3. Datasets for Emotion Classification and Generation**

High-quality, diverse datasets are indispensable for training and validating affective models. The following represent a selection of prominent repositories organized by modality and language support.

**Text Datasets:**
• GoEmotions (Google Research, 2024): One of the most comprehensive English-language emotion classification datasets.
• Emotion-Instructions (Ashish Nehra, 2023): Tailored for instruction-tuned LLMs.
• Amplified Emotions and EmotionalIntelligence-50K: Enriched datasets supporting multi-class emotion detection across different scenarios.

• Sentiment140: Often repurposed for fine-grained emotion classification from tweets.

**Speech Datasets:**
• IEMOCAP, Toronto Emotional Speech Set, and Jvnv-Emotional-Speech-Corpus offer high-fidelity emotional audio data with contextual labeling.
• English_Emotional_Speech_Data_by_Microphone and Urdu-Audio-Emotions increase language diversity in training data.
• AudioSet: Large-scale dataset from Google with labeled audio clips, relevant for emotion detection.

**Visual Emotion Datasets:**
• Visual_Emotional_Analysis (FastJobs, 2023) and COVID-19_Weibo_Emotion demonstrate visual sentiment dynamics in both images and social media posts.
• **FER2013:** Popular dataset with grayscale images, 7 basic emotions.
• **CK+:** Focuses on spontaneous expressions with detailed annotations.
• **AffectNet:** Large-scale dataset (>1M images), 8 emotions, categorical and dimensional.
• **Aff-Wild2:** Multimodal videos of emotions in real-world ("in-the-wild") scenarios.

## Multimodal Datasets:

• CMU-Multimodal (e.g., CMU-MOSI): Benchmark for sentiment analysis with audio-video data
• K-EmoCon: Physiological, audiovisual, and continuous emotion annotations from debates.
• MIB Datasets: Persian language dataset with text, audio, and video.

### 4. Complementary Academic Publications

**Foundational Work:**

• Picard, R. (1997). *Affective Computing*. MIT Press. - Introduced the field.
• Ekman, P. Research on basic universal emotions and facial expressions. - Foundational framework.
• Russell, J. A. (1980). A circumplex model of affect. *Journal of Personality and Social Psychology*. - Dimensional emotion model.
• Ekman, P., & Friesen, W. V. (1978). *Facial Action Coding System (FACS)*. - Detailed facial movement measurement.

**Core Advances and Technical Innovations**
• Cambria, E., Zhang, X., Mao, R., Chen, M., et al. (2024). *SenticNet 8: Fusing Emotion AI and Commonsense AI for Interpretable, Trustworthy, and Explainable Affective Computing*. In Proceedings of the International Conference on Human-Computer Interaction. Springer.
• Nidhishree, M. S., Amutha, S., et al. (2024). *In-Depth Exploration of the Multifaceted Landscape of Emotion AI: From Education and Workplace to Psychiatry*. IEEE IoT Conference Proceedings.
• MDPI. (n.d.). Advances in Facial Expression Recognition: A Survey of Methods, Benchmarks, Models, and Datasets. *Applied Sciences*. https://www.mdpi.com/2078-2489/15/3/135

**Multimodal and Technical Innovations**
• Fang, Y., Diallo, A., Shi, Y., Jumelle, F., & Shi, B. (2025). *End-to-End Facial Expression Detection in Long Videos*. arXiv:2504.07660.
• Lin, Y., Sun, J., Cheng, Z.-Q., Wang, J., et al. (2025). *Why We Feel: Breaking Boundaries in Emotional Reasoning with Multimodal Large Language Models*. arXiv:2504.07521.
• Maharjan, R. S., Romeo, M., & Cangelosi, A. (2025). *Attributes-Aware Visual Emotion Representation Learning*. arXiv:2504.06578.
• Pan, G., Wu, Z., Yang, Y., Wang, X., et al. (2025). *Dynamic Neural Synchrony as a Potential Indicator for Continuous Emotion Arousal*. arXiv:2504.03643.
Prakash, A., & Poulose, A. (2025). *Electroencephalogram-Based Emotion Recognition: A Comparative Analysis of Supervised Machine Learning Algorithms*. Data Science and Management. https://doi.org/10.1016/j.dsm.2024.12.004

**Ethics, Vulnerability, and Social Implications**
• Nagy, J. (2024). *Autism and the Making of Emotion AI: Disability as Resource for Surveillance Capitalism*. New Media & Society.

• Katirai, A. (2024). *Ethical Considerations in Emotion Recognition Technologies: A Review of the Literature.* AI and Ethics. Springer.
• Ali, M., & Ferrucci, P. (2025). *'Zero Human Emotion': AI Anchors and the Normative Repercussions.* Journal of Broadcasting & Electronic Media.
• Frontiers in Psychiatry. (2024). The ethical aspects of integrating sentiment and emotion analysis in chatbots for depression intervention.
• PubMed Central (PMC). (n.d.). An analytical framework for studying attitude towards emotional AI: The three-pronged approach. (Assuming this links to a peer-reviewed journal article)
• PubMed Central (PMC). (n.d.). The Ethics of Emotional Artificial Intelligence: A Mixed Method Analysis. (Assuming this links to a peer-reviewed journal article)
• ACL Anthology. (2023). Pre-trained Speech Processing Models Contain Human-Like Biases that Propagate to Speech Emotion Recognition. *Findings of the Association for Computational Linguistics: EMNLP*.

**Education, Emotions, and AI**
• Yang, L., & Zhao, S. (2024). *AI-Induced Emotions in L2 Education: Exploring EFL Students' Perceived Emotions and Regulation Strategies.* Computers in Human Behavior.
• Yin, H., Wang, C., & Liu, Z. (2024). *Unleashing Pre-Service Language Teachers' Productivity with Generative AI: Emotions, Appraisal, and Coping Strategies.* Computers in Human Behavior.
• Cédric Sarré, Muriel Grosbois, Cédric Brudermann. How does ChatGPT compare with human-generated corrective feedback for grammatical accuracy development in EFL learners' written productions?. EUROCALL 2024, University of Trnava, Aug 2024, Trnava, Slovakia. ⟨hal-04688079⟩
• Frontiers in Psychology. (2024). Integrating artificial intelligence to assess emotions in learning environments: a systematic literature review.

**Applications and Domain-Specific Use Cases**
• Guo, Y., Li, Y., Liu, D., & Xu, S. X. (2024). *Measuring Service Quality Based on Customer Emotion: An Explainable AI Approach.* Decision Support Systems.
• Cho, H. H., Zeng, J. Y., & Tsai, M. Y. (2025). *Efficient Defense Against Adversarial Attacks on Multimodal Emotion AI Models.* IEEE Transactions on Dependable and Secure Computing.
• Phang, J., Lampe, M., Ahmad, L., Agarwal, S., et al. (2025). *Investigating Affective Use and Emotional Well-being on ChatGPT.* arXiv:2504.03888.